\documentclass[aps,reprint,amsmath,amssymb,nofootinbib,superscriptaddress, showpacs,floatfix,prb]{revtex4-1}

\pdfoutput=1

\usepackage{latexsym}
\usepackage{graphicx}
\usepackage{times,psfrag,subfigure}
\usepackage{amsmath}
\usepackage{dcolumn}
\usepackage{bm, bbm}      
\usepackage{color}
\usepackage{latexsym,amsmath,amssymb,bm,euscript}
\newcommand{\beq}{\begin{equation}}
\newcommand{\eeq}{\end{equation}}
\newcommand{\beqarray}{\begin{eqnarray}}
\newcommand{\eeqarray}{\end{eqnarray}}
\newcommand{\Ref}[1]{Ref.~\onlinecite{#1}} 
\newcommand{\eq}[1]{Eq.~(\ref{#1})} 
\newcommand{\fig}[1]{Fig.~\ref{#1}} 
\newcommand{\bsigma}{\mbox{\boldmath$\sigma$}}

\begin{document}

\allowdisplaybreaks

\title{Topologically protected flat zero-energy surface bands in
non-centrosymmetric superconductors} 

\date{\today}

\author{P. M. R. Brydon}
\email{brydon@theory.phy.tu-dresden.de}
\affiliation{Institut f\"ur Theoretische Physik, Technische Universit\"at
  Dresden, D-01062 Dresden, Germany}

\author{Andreas P. Schnyder}
\email{a.schnyder@fkf.mpg.de}
\affiliation{Max-Planck-Institut f\"ur Festk\"orperforschung,
  Heisenbergstrasse 1, D-70569 Stuttgart, Germany} 

\author{Carsten Timm}
\affiliation{Institut f\"ur Theoretische Physik, Technische Universit\"at
  Dresden, D-01062 Dresden, Germany}

\begin{abstract}

Nodal non-centrosymmetric superconductors (NCS) have recently been shown to be
  topologically non-trivial.
An important consequence is the existence of topologically protected
flat zero-energy
surface bands, which are related to the topological characteristics of
the line nodes of the bulk gap via a bulk-boundary correspondence 
  [Schnyder and Ryu, arXiv:1011.1438]. In this 
paper we examine these zero-energy surface bands using a quasiclassical
theory. We determine their spectrum and derive a general condition for their
existence in terms of the sign change of the gap functions. 
A key experimental signature of the zero-energy surface bands is a zero-bias
peak in the tunneling 
conductance, which depends strongly on the surface
orientation. This can be used as a fingerprint of a topologically
non-trivial NCS.


\end{abstract}
\date{\today}



\maketitle

\emph{Introduction}. A key experimental signature of topological insulators
and superconductors is 
the existence of topologically protected zero-energy surface or edge states,
some of which are of Majorana
type.~\cite{hasanKane2010,qiZhangReview2010,ryuNJP10} But zero-energy boundary
modes of topological origin can also occur in gapless topological
systems that exhibit topologically stable Fermi points or in nodal
superconductors with non-trivial topology. In such systems one generically
finds topologically protected dispersionless zero-energy surface
states, i.e., flat bands at the surface. Such flat bands are known to occur at
the
zig-zag and bearded edge in graphene,~\cite{nakada1996,fujita1996} on the (110)
surface of a $d_{x^2-y^2}$-wave
superconductor,~\cite{hu1994,kashiwaya_tanaka00} within vortices in the A phase
of  
${}^3$He,~\cite{VolovikBulkVortex} and in other systems with topologically
protected Dirac points.~\cite{Heikkila2010}

It has recently been realized that nodal non-cen\-tro\-sym\-me\-tric
superconductors
(NCS) are topologically non-trivial states of matter.~\cite{satoPRB06,beriPRB2010,Schnyder2010} An
NCS is realized in a system lacking inversion symmetry, which gives rise to an 
antisymmetric spin-orbit (SO) coupling and consequently to the admixture of
spin-singlet and spin-triplet components in the superconducting state. There
is a steadily growing list of these remarkable materials, most notably
Li$_2$Pd$_x$Pt$_{3-x}$B,~\cite{togano04,badica05} Y$_2$C$_3$,~\cite{amano04} and
the heavy-fermion compounds CePt$_3$Si,~\cite{bauer04}
CeRhSi$_3$,~\cite{kimura05} and CeIrSi$_3$.~\cite{sugitani06} It is predicted
that the topological non-triviality of a nodal NCS leads to topologically
protected zero-energy surface bands, which
only occur within regions of the surface Brillouin zone bounded by the
projected nodal lines of the bulk gap. Since these zero-energy surface bands
give a
singular contribution to the surface density of states, we can expect them to
lead to a zero-bias conductance peak (ZBCP).

It is the aim of this paper to investigate the appearance of zero-energy
surface bands in an NCS using the quasiclassical scattering
theory.~\cite{kashiwaya_tanaka00} This method is ideal for exploring the bound
surface states of unconventional superconductors, and has revealed key aspects
of the surface physics of the cuprate
high-$T_c$ compounds~\cite{hu1994,kashiwaya_tanaka00} and NCS 
systems.~\cite{eschrigIniotakisReview,Tanaka2009,Tanaka2010,Yokoyama2005,Vorontsov2008,Iniotakis2007}
We hence derive the surface-bound-state spectrum and a general
condition for the existence of the zero-energy surface bands in terms of
a sign change of the 
superconducting gap function across the Fermi surface. This condition is
complementary to the topological criterion given in
Ref.~\onlinecite{Schnyder2010}. We then compute the tunneling conductance
between a normal metal and an NCS as a function of both the surface orientation
and the relative magnitude of the spin-singlet and spin-triplet
pairing states. We argue that the strong dependence of the ZBCP on both these
variables provides a powerful diagnostic test 
of the pairing state of an NCS. 


\begin{figure}
  \includegraphics[clip,width=\columnwidth]{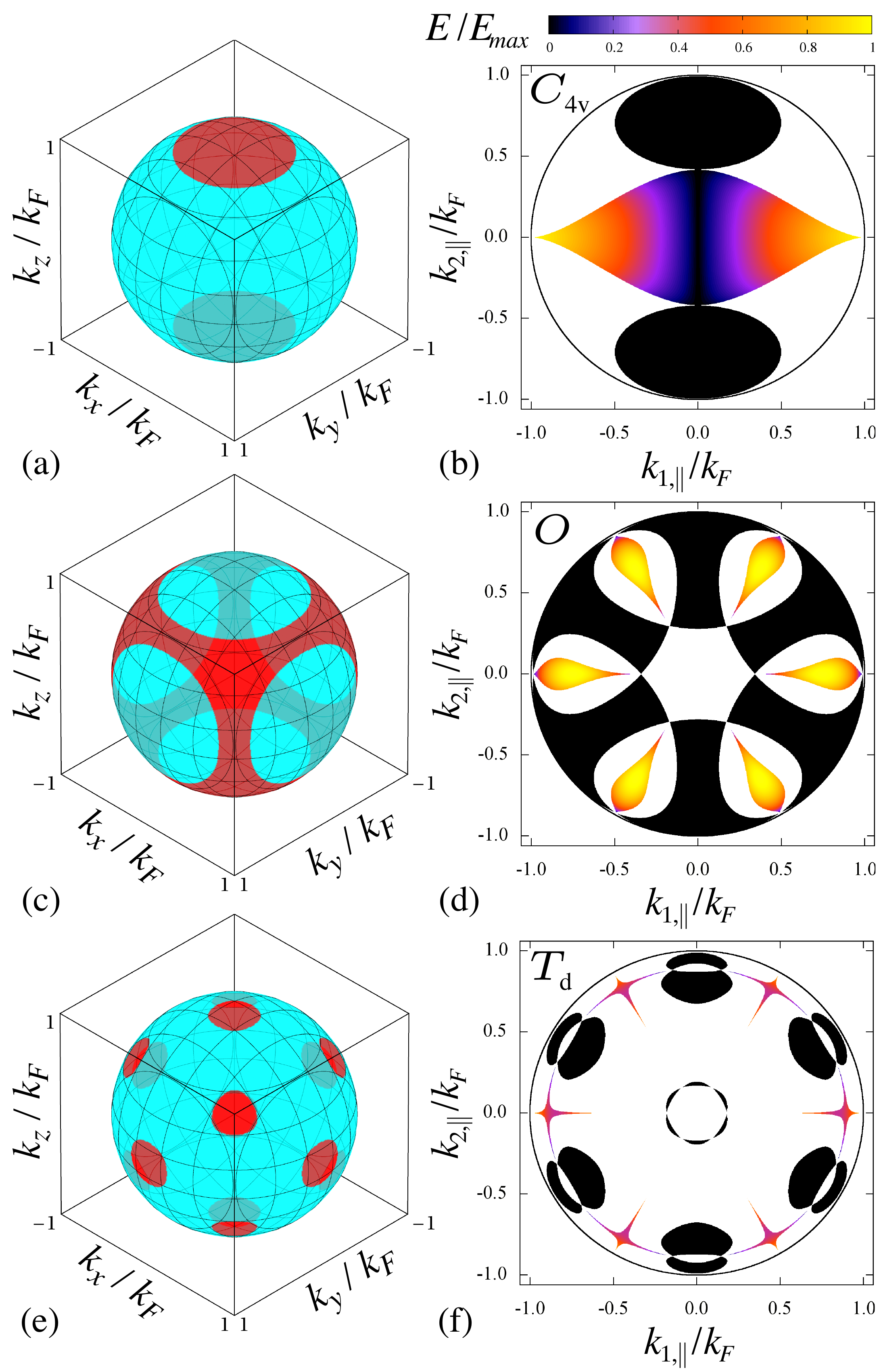}
  \caption{\label{bs_111}(color online) Panels (a), (c) and (e): Variation of
    the sign of $\Delta^{-}_{\bf k}$ over the Fermi surface as seen along
    the $[111]$ direction for (a) the $C_{4v}$ point group with $q=0.5$, (c)
    the $O$ point group with $q=0.35$, and (e) the $T_{d}$ point group with
    $q=0.4$. Dark red indicates $\mbox{sgn}(\Delta^{-}_{\bf k})=1$, whereas
    light blue is $\mbox{sgn}(\Delta^{-}_{\bf k})=-1$. Panels (b), (d) and
    (f): Surface bound states at the $(111)$ 
    face corresponding to the same parameters as in (a), (c) and (e),
    respectively. The color scale of each plot is normalized by the maximum
    bound state energy, (b) $E_{\text max}=0.333\,\Delta_{0}$, (d)
    $E_{\text max}=0.211\,\Delta_{0}$, and (f) $E_{\text max} =
    0.421\,\Delta_{0}$. White
    space indicates the absence of a bound state, and the circle
    is the projection of the Fermi surface, i.e., $|{\bf
      k}_{\parallel}|=k_F$.}
\end{figure}

\emph{Model system}.---We phenomenologically model the NCS as a single-band
system, described by the Bogoliubov-de Gennes (BdG) Hamiltonian 
\begin{eqnarray} \label{ham}
\mathcal{H} ( {\bf k} )
&=&
\begin{pmatrix}
\varepsilon ( {\bf k} )  +  {\bf g} ( {\bf k} ) \cdot \bsigma & 
\Delta ( {\bf k} ) \cr
\Delta^{\dag} ( {\bf k} ) & - \varepsilon ( {\bf k} )  + {\bf g} ( {\bf k}
) \cdot \bsigma^{\ast} \cr
\end{pmatrix},
\end{eqnarray}
where $\varepsilon ( {\bf k} ) $ is the spin-independent part of the band dispersion,
 ${\bf g} ( {\bf k} ) = - {\bf g} ( - {\bf k} ) $ is the  antisymmetric SO
coupling, and 
$ \Delta ( {\bf k} ) = \left( \Delta_s  + {\bf d} ( {\bf k} ) \cdot \bsigma
\right) ( i \sigma_y ) $
is the superconducting gap function.
It is convenient to express the Hamiltonian \eq{ham} in the so-called
helicity basis which diagonalizes the kinetic term, yielding  two helicity
bands with dispersions 
$\xi^{\pm} ( {{\bf k}} ) =  \varepsilon ( {\bf k} ) \pm \left| {\bf g} ( {\bf
  k} ) \right|$.
In the absence of interband pairing the critical temperature is maximized by
taking the 
spin-triplet pairing vector ${\bf d}({\bf k})$ to be aligned with the
polarization vector of the SO coupling ${\bf g} ( {\bf k}
)$,~\cite{frigeriPRL04}
i.e., we parametrize the triplet component of the gap function and the SO
coupling as ${\bf d}({\bf k}) = \Delta_t  {\bf l}_{ {\bf k} }   $  
and ${\bf g} ( {\bf k} ) =  \alpha  {\bf l}_{ {\bf k} } $,
respectively. Hence, the gaps on the two helicity bands are 
 $\Delta^{\pm}_{ {\bf k} } = \Delta_s \pm \Delta_t \left| {\bf l}_{ {\bf k} }
\right|  
 = \Delta_0 ( q \pm  \left| {\bf l}_{{\bf k}} \right| ) / (q+1) $,
 where the parameter $q=\Delta_{s}/\Delta_{t}$ interpolates between
 pure triplet pairing ($q=0$) and pure singlet pairing  ($q \to \infty $).
For simplicity we assume the pairing amplitudes $\Delta_s$ and $\Delta_t$ to
be constant and have positive sign.~\cite{sign} We note that higher-order
angular momentum components of the gap $\Delta_0$ have also been
studied.~\cite{Vorontsov2008,Tanaka2010,yada2011,satoArxi11}

The specific form of the pseudovector ${\bf l}_{ {\bf k} }$ depends on the
symmetries of the non-centrosymmetric crystal.~\cite{samokhin09}
Ignoring the periodic Brillouin-zone structure we employ a small-momentum
expansion
which for the tetragonal point group $C_{4v}$ (relevant for CePt$_3$Si, CeRhSi$_3$, and CeIrSi$_3$) gives, to lowest order, 
the symmetry-allowed form
\begin{subequations}\label{l-vec}
 \begin{eqnarray}
{\bf l}_{ {\bf k} }  =  k_y \hat{{\bf x}} - k_x \hat{{\bf y}}.
 \end{eqnarray}
For the cubic point group  $O$, which is relevant for Li$_2$Pd$_x$Pt$_{3-x}$B, the expansion of ${\bf l}_{ {\bf k} }$
 reads
\begin{eqnarray} \label{l-vecO}
{\bf l}_{ {\bf k} } 
&=& 
k_x \left(1 + g_2[k_y^2 + k_z^2 ] \right) \hat{{\bf x}}
+ k_y\left(1 + g_2[ k_x^2 + k_z^2] \right) \hat{{\bf y}}	
\nonumber\\
&& \qquad
+ k_z \left(1 + g_2[ k_x^2 + k_y^2] \right) \hat{{\bf z}}.	 
\end{eqnarray}
For the spherical Fermi surfaces considered below, the gap $\Delta^{-}_{\bf
  k}$ only has nodes if $g_2 \ne 0$.
Finally, we also consider the tetrahedral point group
$T_d$, experimentally represented by Y$_2$C$_3$, where ${\bf l}_{ {\bf
    k} }$ takes the form   
\begin{eqnarray}
{\bf l}_{ {\bf k} }
 = 
   k_x (k_y^2 - k_z^2) \hat{{\bf x}}
 +   k_y (k_z^2 - k_x^2) \hat{{\bf y}}
 +    k_z (k_x^2 - k_y^2) \hat{{\bf z}}.\quad
\end{eqnarray}
\end{subequations}

\emph{Surface bound states}.---We wish to solve the BdG equation
$\hat{H}\Psi({\bf r}) = E\Psi({\bf r})$ at a surface for states which decay
into the bulk, i.e., which have energies lying within the gap. We define
coordinates parallel (${\bf 
  r}_\parallel$) and perpendicular ($r_\perp$) to the surface such
that the NCS occupies the half-space 
$r_\perp>0$. For simplicity, we ignore the SO splitting of the
bands and assume coincident spherical helical Fermi surfaces with radius
$k_F$.~\cite{eschrigIniotakisReview,Vorontsov2008,Iniotakis2007} We 
have verified that relaxing this approximation does not alter the condition
for the existence of zero-energy surface bands. It is therefore 
convenient to regard the momentum components in the definition of ${\bf 
  l}_{\bf k}$ [\eq{l-vec}] as normalized by $k_F$. If we take $g_2=-1.5$
in~\eq{l-vecO}, then for all point groups we find that $\Delta^{-}_{\bf k}$
has point nodes for $q=0$, $1$, line nodes for $0<q<1$, and is fully
gapped for $q>1$. Examples of the nodal structure of $\Delta^{-}_{\bf
  k}$ are shown in~\fig{bs_111}.

We obtain the following ansatz for the surface bound state wavefunction 
\beq
\Psi({\bf k}_{\parallel},{\bf r}) = \sum_{n=\pm}\sum_{{\bf p}={\bf
    k},\widetilde{\bf k}}a_{n}({\bf p})\psi_{n}({\bf p})e^{-\kappa^{n}_{{\bf p}}r_{\perp}}e^{i{\bf
    p}\cdot{\bf r}} \label{eq:wavefunction} ,
\eeq
where ${\bf k}=({\bf k}_{\parallel},k_{\perp})$ and $\widetilde{\bf
  k}=({\bf k}_\parallel,-k_{\perp})$ are wavevectors with $|{\bf
  k}|=|\widetilde{\bf k}|=k_F$. The momentum component 
parallel to the surface, ${\bf k}_\parallel$, 
is a good quantum number due to translational invariance. 
The positive and negative helicity components $\psi_{\pm} ( {\bf p} )$ are
given by 
\begin{subequations}
\label{eq:components}
\begin{xalignat}{1}
\psi_{+}({\bf p}) 
& = 
 \left(\begin{array}{cccc}
1, & \frac{l^{x}_{\bf p} + i l^{y}_{\bf p}}{|{\bf l}_{\bf p}| + l^{z}_{\bf p}}, 
  & -\frac{l^{x}_{\bf p} + i l^{y}_{\bf p}}{|{\bf l}_{\bf p}| + l^{z}_{\bf p}}\gamma^{+}_{\bf p}, 
  &\gamma^{+}_{\bf p} \end{array}\right)^{T}\, , 
\\
\psi_{-}({\bf p}) 
& =  
\left(\begin{array}{cccc}
\frac{ l^{x}_{\bf p} - i l^{y}_{\bf p}}{|{\bf l}_{\bf p}| + l^{z}_{\bf p}}, 
  & - 1, & \gamma^{-}_{\bf p}, &\frac{ l^{x}_{\bf p} - i l^{y}_{\bf p}}{|{\bf l}_{\bf p}| + l^{z}_{\bf p}}\gamma^{-}_{\bf p}\end{array}\right)^{T}\, ,
\end{xalignat}
\end{subequations}
respectively, with $\gamma^{\pm}_{\bf p} =  (\Delta^{\pm}_{\bf p})^{-1}
[E - i\, \mbox{sgn}(p_{\perp}) (|\Delta^{\pm}_{\bf p}|^2 -
  E^2)^{1/2} ]$.
The wavefunction components decay into the bulk over the inverse
length scale $\kappa^{\pm}_{{\bf p}} =
 m (\hbar^2|k_{\perp}|)^{-1} (|\Delta^{\pm}_{\bf p}|^2 -
  E^2)^{1/2}$, where $m$ is the effective mass. 

A bound state occurs if the coefficients $a_{n}({\bf p})$
in~\eq{eq:wavefunction} can be chosen so that the wavefunction
vanishes at the surface. After some algebra, this yields the following
condition for surface bound state formation,
\beqarray \label{ABSCondition}    
0 
& = & 
    (\gamma^{+}_{\widetilde{\bf k}} - \gamma^{-}_{\bf k})
    (\gamma^{-}_{\widetilde{ \bf k}} - \gamma^{+}_{\bf k})
    (|{\bf l}_{\bf k}| |{\bf l}_{\widetilde{\bf k}}| - {\bf l}_{\bf k}\cdot{\bf  l}_{\widetilde{\bf k}}) 
\notag \\
& & 
+ (\gamma^{+}_{\widetilde{\bf k}} - \gamma^{+}_{\bf k})
    (\gamma^{-}_{\widetilde{\bf k}} - \gamma^{-}_{\bf k})
    (|{\bf l}_{\bf k}||{\bf l}_{\widetilde{\bf k }}| + {\bf l}_{\bf k}\cdot{\bf l}_{\widetilde{\bf k}}). 
\eeqarray
Setting $E=0$ in~\eq{ABSCondition} and observing that 
$\left. \gamma^{\pm}_{\bf p} \right|_{E=0} = -i \mbox{sgn}( p_{\perp} )
\mbox{sgn} ( \Delta^{\pm}_{{\bf p}} )$,
we find that~\eq{ABSCondition} has a non-trivial zero-energy solution 
whenever (i) $\mbox{sgn} ( \Delta^{-}_{{\bf k}} ) =
\mbox{sgn} (\Delta^{-}_{\widetilde{\bf k}} ) = -1$ and ${\bf l}_{\bf k}\cdot{\bf l}_{\widetilde{\bf k}} = -|{\bf l}_{\bf
  k}||{\bf l}_{\widetilde{\bf k}}|$ or (ii) $\mbox{sgn} ( \Delta^{-}_{{\bf k}} ) =
-\mbox{sgn} ( \Delta^{-}_{\widetilde{\bf k}} )$. 
The latter condition 
corresponds to the topologically protected zero-energy surface bands
found in~\Ref{Schnyder2010}, which occur within a finite region of the surface
Brillouin zone bounded by the projected nodes of the bulk gap. 
It is interesting that this condition is
quite similar to the one for zero-energy states at surfaces of unconventional
{\it centrosymmetric} 
superconductors, i.e., the sign of the gap must reverse between ${\bf
  k}$ and $\widetilde{\bf k}$. It is therefore somewhat surprising that the
relative sign between the gaps on \emph{different} helicity bands at these
  wavevectors is irrelevant, despite the mixing of the helicity components in
the edge-state wavefunction.

In Figs.~\ref{bs_111}(b), (d), and (f) we plot the dispersion of the bound
states appearing 
at the (111) surface of an NCS with point group $C_{4v}$, $O$, and $T_d$,
respectively.  
For all three point groups we find zero-energy surface bands for $0<q<1$,
i.e., there are two-dimensional regions in the surface Brillouin zone
where the energy of the bound states vanishes [black regions in
Figs.~\ref{bs_111}(b), (d), and (f)]. As can be seen by comparison with
Figs.~\ref{bs_111}(a), (c), and (e), these occur where
the sign of $\Delta^{-}_{\bf k}$ on the forward-facing half of the Fermi
surface is different to that on the backward-facing half.
This is a manifestation of the bulk-boundary correspondence found
in~\Ref{Schnyder2010} which relates the topologically stable nodal lines of
the bulk gap to the zero-energy surface bands. 
A quantized topological invariant is associated with the line nodes, which
protects them and hence also the zero-energy surface bands from sufficiently
small symmetry preserving perturbations.
Large perturbations which remove 
the nodal rings (e.g., increasing the singlet-to-triplet ratio $q$ past $1$)
therefore also destroy the associated zero-energy surface bands.

The zero-energy surface bands
are in general accompanied by dispersing modes [colored
regions in Figs.~\ref{bs_111} (b), (d), and (f)], which occur in regions of the
surface 
Brillouin zone where $\mbox{sgn} ( \Delta^{-}_{{\bf k}} ) =
\mbox{sgn} (\Delta^{-}_{\widetilde{\bf k}} ) = -1$. The coexistence of
dispersive and non-dispersive surface states give rise to intricate
bound-state spectra. In the case of the $C_{4v}$ point group
[\fig{bs_111}(a)] the linearly dispersing states at 
sufficiently small $|k_{2,\parallel}|$ arise from the same mechanism as the
states found along the (100) direction
in~\Ref{Iniotakis2007}. Note the line of zero-energy states at
$k_{1,\parallel}=0$ connecting the two flat bands.


\begin{figure}
  \includegraphics[clip,width=1.0\columnwidth]{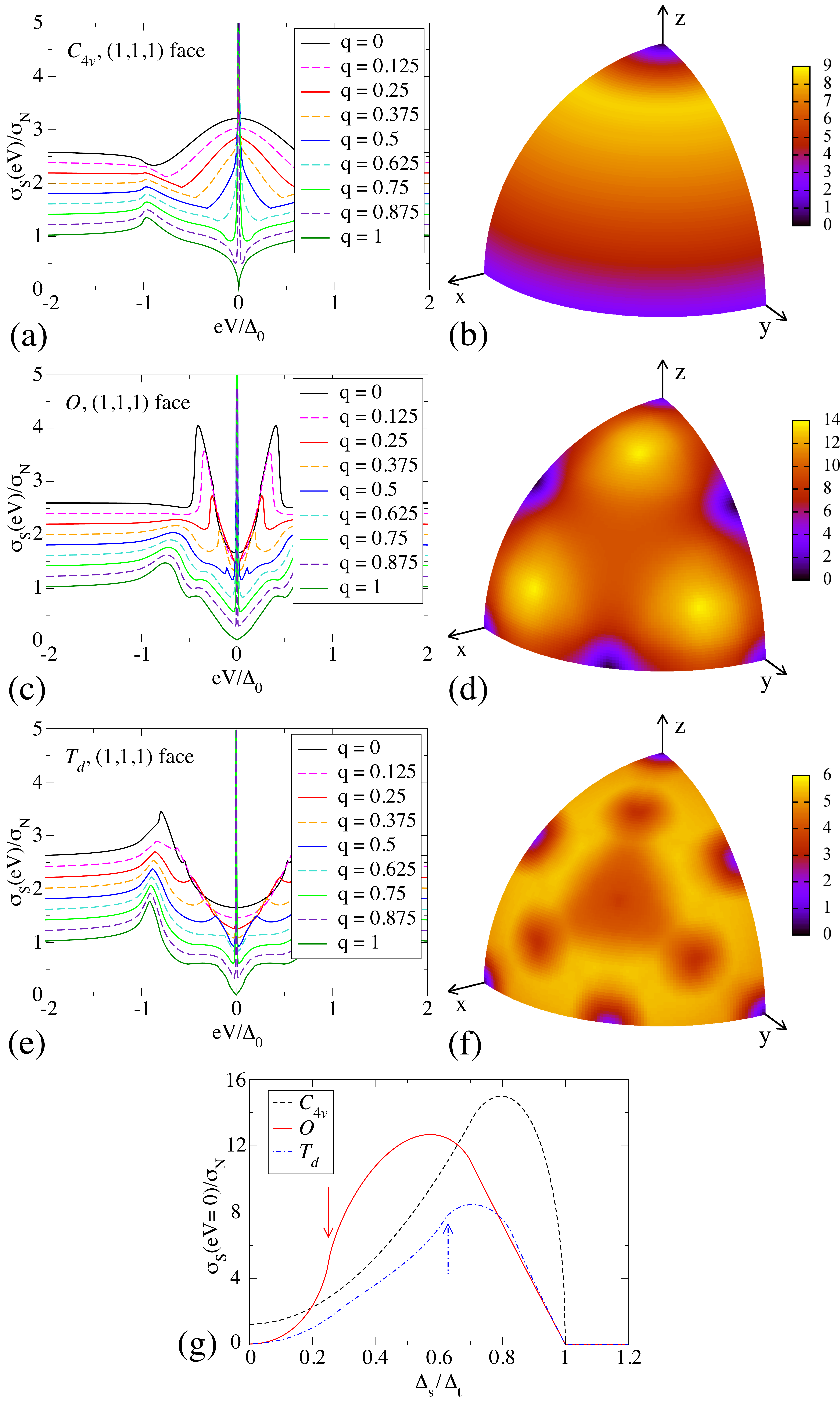}
  \caption{\label{conductance}
  (color online) Panels (a), (c) and (e): Normalized
    conductance spectra at the $(111)$ face for an NCS with point group (a)
    $C_{4v}$, (c) 
    $O$, and (r) $T_d$ and various values of $0\leq
    q=\Delta_{s}/\Delta_{t}\leq1$. In all panels we take $Z=3$ and
    $T=0$\,K. Curves are vertically shifted by multiples of $0.2$. 
      Panels(b), 
    (d) and (f): ZBCP as a function of surface
    orientation for an NCS with point group (b) $C_{4v}$ with $q=0.5$, (d) $O$
    with $q=0.35$, and (f) $T_d$ with $q=0.4$. The color at each point on the
    sphere indicates the height of the ZBCP for the corresponding normal
    vector. Panel (g):
    Variation of ZBCP height at the $(111)$ face as a function of
    $q$.}
\end{figure}

\emph{Tunneling conductance}.---Although the surface bound states do
not form at an interface with a 
normal metal, for a low-transparency barrier between a metal and an NCS, the
physical mechanism discussed above leads to the formation of interface-resonance
states. For a surface which displays zero-energy bands, it is well-known
that the corresponding tunnel
junction will show a sharp ZBCP in the low-temperature tunneling
conductance. As in the case of the $d$-wave gap in the high-$T_c$ 
cuprates,~\cite{kashiwaya_tanaka00,Wei1998} a direction-dependent
ZBCP in an NCS due to the topologically protected
zero-energy
surface bands would be a key experimental signature of the pairing state.

The zero-temperature charge conductance $\sigma_{S}(eV)$ for tunneling
into the NCS is a 
generalization of the usual Blonder-Tinkham-Klapwijk
formula~\cite{Blonder1982,Yokoyama2005,eschrigIniotakisReview,Iniotakis2007} 
\beq
\sigma_{S}(eV) = \sum_{{\bf k}_\parallel}\left\{1 +
  \frac{1}{2}\sum_{\sigma,\sigma'}\left[\left|a_{{\bf
        k}_\parallel}^{\sigma,\sigma'}\right|^2 - \left|b_{{\bf k}_\parallel}^{\sigma,\sigma'}
        \right|^2\right]\right\},
\eeq
where $a_{{\bf k}_\parallel}^{\sigma,\sigma'}$ and $b_{{\bf
    k}_\parallel}^{\sigma,\sigma'}$ are the Andreev and
normal reflection for electron injection into the NCS, respectively. These
coefficients are determined by solving the BdG equation for the
junction at energy $E=eV$, with the wavefunction ansatz $\Psi_{\sigma}({\bf
  k}_\parallel,{\bf r}) = \Theta(-r_\perp)\psi^{<}_{{\bf
  k}_\parallel,\sigma}({\bf r}) + \Theta(r_\perp)\psi^{>}_{{\bf
  k}_\parallel,\sigma}({\bf r})$ where
\beqarray
\psi^{<}_{{\bf k}_\parallel,\sigma}({\bf r}) & = & \psi_{e,\sigma}e^{i{\bf
    k}\cdot{\bf r}} + \sum_{\sigma'=\uparrow,\downarrow}\big[a_{{\bf
      k}_\parallel}^{\sigma,\sigma'}\psi_{h,\sigma'}e^{i{\bf
      k}\cdot{\bf r}} \notag \\
&& {}+ b_{{\bf
      k}_\parallel}^{\sigma,\sigma'}\psi_{e,\sigma'}e^{i\widetilde{\bf
      k}\cdot{\bf r}}\big]\,, \notag \\
\psi^{>}_{{\bf k}_\parallel,\sigma}({\bf r}) & = & \sum_{n=\pm}\big[c_{{\bf k}_\parallel}^{\sigma,n}\psi_{n}({\bf k})e^{i{\bf
    k}\cdot{\bf r}} + d_{{\bf k}_\parallel}^{\sigma,n}\psi_{n}(\widetilde{\bf
    k})e^{i\widetilde{\bf k}\cdot{\bf r}}\big]\,.
\label{eq:scattering}
\eeqarray
$\psi_{e,\sigma} = \frac{1}{2}(1+\sigma,1-\sigma,0,0)^{T}$ and
$\psi_{h,\sigma} = \frac{1}{2}(0,0,1+\sigma,1-\sigma)^{T}$ are
the electron and hole spinors in the normal metal,
respectively. We adopt the assumption that the bias energy is small compared 
to the Fermi energy so that the wavevectors in~\eq{eq:scattering} are
well-approximated to have magnitude $k_F$.~\cite{Yokoyama2005,Tanaka2009} The
insulating
barrier at $r_\perp=0$ is modeled as a $\delta$-function of height $U$. The
coefficients in~\eq{eq:scattering} are then chosen such that 
$\Psi_\sigma$ is continuous at the interface, i.e.,
$\Psi_\sigma({\bf k}_\parallel,{\bf r})|_{r_\perp = 0^{-}} = \Psi_\sigma({\bf
  k}_\parallel,{\bf r})|_{r_\perp=0^{+}}$, 
while the derivative obeys ${\partial_{r_\perp}}\Psi_\sigma({\bf
  k}_\parallel,{\bf r})|_{r_\perp=0^{+}}-{\partial_{r_\perp}}\Psi_\sigma({\bf
  k}_\parallel,{\bf r})|_{r_\perp=0^-} = 2Z
  \Psi_\sigma({\bf k}_\parallel,{\bf r})|_{r_\perp=0}$,
where $Z=mU/\hbar^2$ with $m$ is the effective mass, assumed the same in the
NCS and the metal. 

In Fig.~\ref{conductance}(a), (c), and (e) we show the conductance spectra at
the
(111) face for the $C_{4v}$, $O$, and $T_d$ point groups, respectively. The
spectra are normalized by the normal-state conductance $\sigma_{N}=\sum_{{\bf
    k}_\parallel}|{\bf k}_\parallel|^2/(Z^2 + |{\bf k}_\parallel|^2)$.
For $0<q<1$ we find a
sharp ZBCP, signaling the existence of the zero-energy
surface bands. For the most part, this peak is well-separated from the
contributions of dispersing states or the edges of the bulk gap. 
A notable
exception is the $C_{4v}$ case [\fig{conductance}(a)], where for small $q$ we
find the ZBCP superimposed on a wide dome-like feature
due to the dispersing surface modes.~\cite{eschrigIniotakisReview,Iniotakis2007}
As shown
in~\fig{conductance}(g), the ZBCP
vanishes as $q \rightarrow 1$ and is absent for $q>1$, consistent with the
topologically trivial gapped state which forms when the singlet pairing
dominates. We note that in the case of the $O$ and $T_{d}$ point groups, the
ZBCP height shows kinks as a function of $q$, marked by the arrows
in~\fig{conductance}(g). At these values of $q$ there is a Lifshitz-type
transition in the BdG spectrum at which the nodal rings touch
each other and then reconnect.

Finally, we consider the direction dependence of the ZBCP
[\fig{conductance}(b), (d), and (f)]. The strong variation in 
the conductance with the surface orientation reflects the changing
projection of 
the bulk nodal lines onto the surface Brillouin zone.
Note that for all point groups the ZBCP is absent along
the crystal axes. This can be exploited as a test of the pairing state in an
NCS: For example, the observation of a ZBCP along the
(111) direction, but its absence along the (100) direction, would lend strong
support to a model of an NCS pairing state with a dominant triplet component.

\emph{Conclusions}.---In this paper we have used quasiclassical scattering
theory to study the appearance of topologically protected zero-energy
bands at the surface of non-centrosymmetric superconductors for three
experimentally relevant choices of the point group. We have derived a
general condition for the existence of these states, which is
consistent with the bulk-boundary correspondence found
in~\Ref{Schnyder2010}. The surface-bound-state spectrum has been computed and
shown to allow the coexistence of zero-energy flat bands with dispersing
states. We have also calculated the tunneling conductance, where the presence
of these zero-energy states manifests itself as a ZBCP. The
ZBCP displays a strong dependence on the surface orientation and in
particular vanishes along the crystal axes. We propose that this dependance can
be exploited as a test of the orbital and spin pairing symmetries in these
materials. 


Prospects for further work are promising. For example, the large degeneracy of
the zero-energy surface bands may be expected to lead to instabilities towards
symmetry-broken states in the presence of interactions. The possible
coexistence of time-reversal-symmetry-breaking and
time-reversal-symmetry-preserving order parameters near the surface of an NCS
is particularly tantalizing. 

\emph{Acknowledgements}.---The authors thank A.\ Avella, S.\ Ryu, and
M.\ Sigrist for useful discussions.


\begin{thebibliography}{99}
\bibitem{sign}The relative signs of $\Delta_{s}$ and $\Delta_{t}$ determine
  the presence of nodes in either the positive or negative helicity
  gaps. In the following we regard the SO splitting of the bands as
  insignificant, allowing us to fix the signs without loss of generality.
\end{thebibliography}
\end{document}